\documentclass[12pt]{article}  
\usepackage{cite}
\usepackage{epsfig}
\usepackage{graphicx}
\usepackage{amsmath}
\usepackage{amssymb}
\usepackage{ulem}
\usepackage{mdwlist}          
\usepackage{color}              

\usepackage{a41}
\usepackage{color}
\usepackage[rflt]{floatflt}
\usepackage{float}
\usepackage{slashed}

\usepackage{array}

\usepackage[english]{babel}
\usepackage[latin1]{inputenc}
\usepackage[T1]{fontenc}
\usepackage{ae}

\usepackage{url}

\usepackage{amsmath, amsthm, amssymb}
\newtheorem{thm}{Theorem}[section]

\newtheorem{definition}[thm]{Definition}

 \newcommand{\GeV}{\mathrm{GeV}}
\newcommand{\Mvec}{{\rm\bf M}}

\usepackage{rotating}

\usepackage{graphicx}
\newcounter{mmacnt}
\def\restartmma{\setcounter{mmacnt}{0}}
\restartmma \catcode`|=\active
\def|#1|{\mathrm{#1}}
\catcode`|=12
\newenvironment{mma}{
 \par\smallskip
 \catcode`|=\active
 \parskip=0pt\parindent=0pt % locally
 \small
 \def\In##1\\{%
   \def\linebreak{\hfill\break\null\qquad}%
   \refstepcounter{mmacnt}
   \hangindent=2.5em\hangafter=0
   \leavevmode
   \llap{\tiny\sffamily In[\arabic{mmacnt}]:=\kern.5em}%
   \mathversion{bold}\footnotesize$\displaystyle##1$\normalsize
   \mathversion{normal}\par
 }%
 \def\Print##1\\{%
   \def\linebreak{\hfill\break}%
   \hangindent=2.5em\hangafter=0
   \leavevmode ##1\par}%
 \def\Out##1\\{%
   \def\linebreak{$\hfill\break\null\hfill$}%
   \kern\abovedisplayskip\par
   \hangindent=2.5em\hangafter=0
   \leavevmode
   \llap{\tiny\sffamily Out[\arabic{mmacnt}]=\kern.5em}
   \footnotesize$\displaystyle##1$\normalsize\hfill\null\par
   \kern\belowdisplayskip
 }%
 \def\Warning##1##2\\{%
   \def\linebreak{\hfill\break}%
   \hangindent=2.5em\hangafter=0
   \leavevmode
   {\scriptsize##1 : ##2}\par}%
}{%
 \par\smallskip
}

\usepackage{color}

\newenvironment{fshaded}{%
\MakeFramed {\FrameRestore}
}%
{\endMakeFramed}

\allowdisplaybreaks[4]

\begin{document}
\setlength{\baselineskip}{0.515cm}
\sloppy
\thispagestyle{empty}
\begin{flushleft}
DESY 17--187
%\hfill {\tt arXiv:1804.xxxxx[hep-ph]}
\\
DO--TH 17/30\\
April 2018\\
\end{flushleft}

\mbox{}
\vspace*{\fill}
\begin{center}

{\LARGE\bf The Variable Flavor Number Scheme at} 

\vspace*{3mm} 
{\LARGE\bf Next-to-Leading Order}

\vspace{3cm}
\large
J.~Bl\"umlein$^a$, 
A.~De Freitas$^a$, 
C.~Schneider$^b$, and
K.~Sch\"onwald$^a$

\vspace{1.cm}
\normalsize
{\it  $^a$ Deutsches Elektronen--Synchrotron, DESY,}\\
{\it  Platanenallee 6, D-15738 Zeuthen, Germany}
\\

\vspace*{3mm}
{\it $^b$~Research Institute for Symbolic Computation (RISC),\\
                          Johannes Kepler University, Altenbergerstra\ss{}e 69,
                          A--4040 Linz, Austria}\\

%%\today

\end{center}
\normalsize
\vspace{\fill}
\begin{abstract}
\noindent
We present the matching relations of the variable flavor number scheme at next-to-leading order, which 
are of importance to define heavy quark partonic distributions for the use at high energy colliders such 
as Tevatron and the LHC. The consideration of the two-mass effects due to both charm and bottom quarks, 
having rather similar masses, are important. These effects have not been considered in previous investigations.
Numerical results are presented for a wide range of scales. We also present the corresponding contributions
to the structure function $F_2(x,Q^2)$.
\end{abstract}

\vspace*{\fill}
\noindent
%\numberwithin{equation}{section}

\newpage

\vspace*{1mm}
\noindent
In the variable flavor number scheme (VFNS), matching conditions are considered between parton distribution 
functions (PDFs) at $N_F$ massless flavors and those at $N_F + k$ flavors (with usually $k=1,2$), at
high factorization and renormalization scales $\mu^2$. This allows to introduce heavy quark parton distribution 
functions, which are related to the quark-singlet ($\Sigma$) and gluon ($G$) distributions via the universal 
massive operator matrix elements (OMEs) $A_{ij}^{(k)}(\mu^2,m_c^2,m_b^2)$. Likewise, the flavor non-singlet, 
singlet and gluon distribution functions receive corresponding QCD-corrections. In this paper we will work 
in the $\overline{\sf MS}$-scheme in QCD, defining the heavy quark masses first in the on-shell scheme and later 
also transforming to the $\overline{\sf MS}$-scheme. The VFNS for $k = 1$ has been discussed in 
Ref.~\cite{Buza:1996wv} 
at NLO and at NNLO in \cite{Bierenbaum:2009mv} and including the two-mass effects in Ref.~\cite{Ablinger:2017err} 
to NNLO. 

In the past the usual approach has been to deal with a single heavy quark at a time. However,
the charm and bottom quarks have rather similar masses with $m_c^2/m_b^2 \sim 1/10$ for their pole 
or $\overline{\sf MS}$ masses at NLO and NNLO, which makes it difficult to assume $m_c^2 \ll m_b^2$, 
i.e. to consider the charm mass at $\mu = m_b$ massless. On the other hand, it is perfectly possible
to decouple both quarks simultaneously and consider their effect at high scales $\mu \gg m_c, m_b$. 
In this note, we will describe the VFNS in this more general case at next-to-leading order.

The parton distributions for $N_F + 2$ flavors are related to those at $N_F$ flavors by the following 
relations for the {\it number} densities in Mellin-$N$ space
%------------------------------------------------------------------------------------------------------
\begin{eqnarray}
\label{eq:FI}
f_{{\sf NS},i}(N_F+2,\mu^2) &=& \Biggl\{1 + a_s^2(\mu^2) \left[
 A_{qq,Q}^{{\sf NS},(2,c)}
+A_{qq,Q}^{{\sf NS},(2,b)}\right] \Biggr\} f_{{\sf NS},i}(N_F,\mu^2),
\\
%-------
\label{eq:FIS}
\Sigma(N_F+2,\mu^2) &=& \Biggl\{1 + a_s^2(\mu^2)\Bigl[
 A_{qq,Q}^{{\sf NS},(2,c)} + A_{qq,Q}^{{\sf PS},(2,c)}
+A_{qq,Q}^{{\sf NS},(2,b)} + A_{qq,Q}^{{\sf PS},(2,b)} \Bigr] \Biggr\}
\Sigma(N_F,\mu^2)
\nonumber\\ &&
\hspace*{-3mm}
+ \Biggl\{a_s(\mu^2) \Bigl[
 A_{Qg}^{(1,c)}
+A_{Qg}^{(1,b)}\Bigr] 
+a_s^2(\mu^2) 
\Bigl[
 A_{Qg}^{(2,c)}
+A_{Qg}^{(2,b)} + A_{Qg}^{(2,cb)}\Bigr]\Biggr\} G(N_F,\mu^2),
\\
%-----
\label{eq:GNF2}
G(N_F+2,\mu^2) &=& 
 \Biggl\{1 + a_s(\mu^2) \Bigl[
 A_{gg,Q}^{(1,c)}
+A_{gg,Q}^{(1,b)}\Bigr] 
+a_s^2(\mu^2) 
\Bigl[
 A_{gg,Q}^{(2,c)}
+A_{gg,Q}^{(2,b)} + A_{gg,Q}^{(2,cb)}\Bigr]\Biggr\} G(N_F,\mu^2)
\nonumber\\
&&
+a_s^2(\mu^2)\Bigl[
  A_{gq,Q}^{(2,c)}
+ A_{gq,Q}^{(2,b)} \Bigr] \Sigma(N_F,\mu^2), \nonumber\\ &&
\\
%------
\label{eq:cc}
\lefteqn{\hspace*{-2.5cm}\Bigl[f_c+f_{\bar{c}}\Bigr](N_F+2,\mu^2) = a_s^2(\mu^2)
  A_{Qq}^{{\sf PS},(2,c)}
 \Sigma(N_F,\mu^2) 
} 
\nonumber\\
&&~~~~~~
+ \Biggl\{a_s(\mu^2) 
A_{Qg}^{(1,c)}
+a_s^2(\mu^2) 
\Bigl[
 A_{Qg}^{(2,c)}
+ \frac{1}{2} A_{Qg}^{(2,cb)}\Bigr]\Biggr\} G(N_F,\mu^2),
\\
%------
\label{eq:ma2VNS}
\lefteqn{\hspace*{-2.5cm}\Bigl[f_b+f_{\bar{b}}\Bigr](N_F+2,\mu^2) = a_s^2(\mu^2)
  A_{Qq}^{{\sf PS},(2,b)}
 \Sigma(N_F,\mu^2) 
} 
\nonumber\\
&&~~~~~~
+ \Biggl\{a_s(\mu^2) 
A_{Qg}^{(1,b)}
+a_s^2(\mu^2) 
\Bigl[
 A_{Qg}^{(2,b)}
+ \frac{1}{2} A_{Qg}^{(2,cb)}\Bigr]\Biggr\} G(N_F,\mu^2)~.
\end{eqnarray}
%------------------------------------------------------------------------------------------------------
Here $a_s = \alpha_s/(4\pi) = g_s^2/(4\pi)^2$ denotes the strong coupling constant. The quark non-singlet and 
singlet distributions are defined by\footnote{Actually, one 
should subtract from (\ref{eq:NS1}) the term $\Sigma/N_F$. However, as the functional 
relations are the same, we follow the convention suggested in Ref.~\cite{Buza:1995ie}.} 
%------------------------------------------------------------------------------------------------------
\begin{eqnarray}
\label{eq:NS1}
f_{{\sf NS},i}(N_F,\mu^2) &=& q_i(\mu^2) + \bar{q}_i(\mu^2),
\\
\Sigma(N_F,\mu^2) &=& \sum_{i=1}^{N_F} \left[q_i(\mu^2) + \bar{q}_i(\mu^2)\right].
\label{eq:LA}
\end{eqnarray}
%------------------------------------------------------------------------------------------------------
The OMEs $A_{ij}^{(k,Q)}$ and $A_{ij}^{(k,Q_1 Q_2)}$ depend on $\mu^2/m_Q^2$ and 
$\mu^2/m_{Q_i}^2$ logarithmically. Eqs.~(\ref{eq:FI}--\ref{eq:ma2VNS}) describe the corresponding heavy
flavor contributions at $N_F+2$ flavors in fixed order perturbation theory. Here
we have dropped the dependence on $N$ of the contributing functions.

In $x$-space the convolutions are given by
%------------------------------------------------------------------------------------------------------
\begin{eqnarray}
([f]_+ \otimes g)(x) &=&  \int_x^1 dz \frac{1-z}{z} f(z) g\left(\frac{x}{z}\right) - g(x) \int_0^x dz f(z)
%%%\int_x^1 dz f\left(\frac{x}{y}\right) \left[ \frac{1}{y} g(y) - 
%%%\frac{x}{y^2} g(x) \right] - g(x) \int_0^x dz f(z), \\
\\
%---
(h \otimes g)(x) &=& \int_x^1 \frac{dz}{z} h(z) g\left(\frac{x}{z}\right),~~~~(\delta(1-x) \otimes g)(x) = 
g(x),
\end{eqnarray}
%------------------------------------------------------------------------------------------------------
for a +-distribution $[f]_+$, regular functions $g$ and $h$, and the convolution with the 
$\delta(1-x)$-distribution. Again,
we consider the case of number densities for $g(x)$ here. The +-distribution has the Mellin transform
%------------------------------------------------------------------------------------------------------
\begin{eqnarray}
\Mvec[[f(x)]_+](N) = \int_0^1 dx \left[x^{N-1}-1\right] f(x)~,
\end{eqnarray}
%------------------------------------------------------------------------------------------------------
and the Mellin transform in general obeys
%------------------------------------------------------------------------------------------------------
\begin{eqnarray}
\Mvec[(h \otimes g)(x)](N) = 
\Mvec[h(x)](N) \cdot \Mvec[g(x)](N)~.
\end{eqnarray}
%------------------------------------------------------------------------------------------------------

The flavor non-singlet distributions are not effected by two-mass terms at NLO, but first at NNLO, 
cf.~\cite{Ablinger:2014vwa,Ablinger:2017err}. The OMEs to NLO in Eqs.~(\ref{eq:FI}--\ref{eq:ma2VNS})
have been calculated in Refs.~\cite{Buza:1995ie,Buza:1996wv,Bierenbaum:2007qe,Bierenbaum:2009zt,Behring:2014eya} 
in the equal mass case. At NNLO the OMEs have been computed for a series of moments in \cite{Bierenbaum:2009mv} 
and for a part of the OMEs for general moments $N$ in \cite{Ablinger:2010ty,
Ablinger:2014vwa,Ablinger:2014nga,Ablinger:2014lka,AGG,Ablinger:2016kgz,Ablinger:2017ptf,Ablinger:2015tua,Behring:2014eya} 
in the equal mass 
case. In the unequal mass case at NNLO the moments $N = 2,4,6$ of all OMEs were calculated in terms of an 
expansion in the mass ratio in \cite{Ablinger:2017err} and a 
part of the general $x$ corrections have been computed in \cite{Ablinger:2017err,Ablinger:2017xml,AGG2} already.

The unequal mass corrections at NLO in Eqs.~(\ref{eq:FI}--\ref{eq:ma2VNS}) were calculated in 
Ref.~\cite{Ablinger:2017err}.
They are given by
%------------------------------------------------------------------------------------------------------
\begin{eqnarray}
\label{eq:tm1}
A_{Qg}^{(2,cb)}   &=& 
-\beta_{0,Q} \hat{\gamma}_{qg}^{(0)}  
\ln\left(\frac{\mu^2}{m_c^2}\right)
\ln\left(\frac{\mu^2}{m_b^2}\right),
\\
\label{eq:tm2}
A_{gg,Q}^{(2,cb)}   &=& 
2 \beta_{0,Q}^2 
\ln\left(\frac{\mu^2}{m_c^2}\right)
\ln\left(\frac{\mu^2}{m_b^2}\right),
\end{eqnarray}
%------------------------------------------------------------------------------------------------------
where $\beta_{0,Q} = -(4/3) T_F$ and $T_F = 1/2$ and 
%------------------------------------------------------------------------------------------------------
\begin{eqnarray}
\hat{\gamma}_{qg}^{(0)} = -8 T_F \frac{N^2 + N +2}{N(N+1)(N+2)},
\end{eqnarray}
%------------------------------------------------------------------------------------------------------
denotes the leading order splitting function for the process $g \rightarrow q$~\footnote{As very well known,
splitting functions, to all orders in the coupling constant, are universal and do especially not contain 
power corrections $m^2/Q^2$ stemming e.g. from phase space corrections.}.
The following sum rule has to be obeyed due to energy-momentum conservation, cf.~\cite{Bierenbaum:2009mv},
%------------------------------------------------------------------------------------------------------
\begin{eqnarray}
\label{eq:SR}
A_{Qg}(N=2) + A_{qg,Q}(N=2) + A_{gg,Q}(N=2) = 1.
\end{eqnarray}
%------------------------------------------------------------------------------------------------------
The OME $A_{qg,Q}$ contributes from 3-loop order onwards only and has two heavy quark contributions
only beginning at 4-loop order. 
The equal mass terms are already known to obey Eq.~(\ref{eq:SR}) up to $O(a_s^3)$, 
\cite{Bierenbaum:2009mv}. The NLO
mass contributions add up to zero for $N=2$ in accordance with Eq.~(\ref{eq:SR}).

To illustrate the numerical effect of the NLO 2-mass terms on these distributions we consider the ratio
%------------------------------------------------------------------------------------------------------
\begin{eqnarray}
\alpha \frac{a_s^2(\mu^2) A_{ig}^{(2,cb)} G(N_F,\mu^2)}{\Phi(N_F+2,\mu^2)},
\label{eq:RATIO}
\end{eqnarray}
%------------------------------------------------------------------------------------------------------
for $\Phi = \Sigma, G, (\alpha=1); [f_c+f_{\bar{c}}], [f_b+f_{\bar{b}}], (\alpha = 1/2).$ In the case of the 
heavy flavor distributions, the effect is largest because it is of $O(a_s)$.
A first simple estimate yields
%------------------------------------------------------------------------------------------------------
\begin{eqnarray}
\label{eq:EST}
\frac{\Bigl[f_c+f_{\bar{c}}\Bigr]^{\rm two mass}(N_F+2,\mu^2)}
     {\Bigl[f_c+f_{\bar{c}}\Bigr]^{\rm all}(N_F+2,\mu^2)} \approx a_s \left[\beta_{0,Q}  
\ln\left(\frac{\mu^2}{m_b^2}\right)
+ O(a_s)\right],
\end{eqnarray}
%------------------------------------------------------------------------------------------------------
and similar for $[f_b+f_{\bar{b}}]$ by exchanging $c \leftrightarrow b$. Here the leading term 
does not depend on the parton distributions in Mellin space.

For all contributions to the OMEs but $A_{Qg}^{(2)}$ and $A_{gg,Q}^{(2)}$ the same relation is obtained in the 
$\overline{\sf MS}$ and on-shell scheme to $O(a_s^2)$ for mass renormalization. The transition relations for 
$A_{Qg}^{(2)}$ and $A_{gg,Q}^{(2)}$ for the single mass terms read
%------------------------------------------------------------------------------------------------------
\begin{eqnarray}
A_{Qg}^{\overline{\sf MS}(2)}(\overline{m})   &=& A_{Qg}^{{\sf OS}(2)}(m = \overline{m}) +
4 C_F \hat{\gamma}_{qg}^{(0)}\left[1 + \frac{3}{4} \ln\left(\frac{\mu^2}{\overline{m}^2}\right) \right]
\\
A_{gg,Q}^{\overline{\sf MS}(2)}(\overline{m}) &=& A_{gg,Q}^{{\sf OS}(2)}(m = \overline{m}) - 
8  C_F \beta_{0,Q}\left[1 + \frac{3}{4} \ln\left(\frac{\mu^2}{\overline{m}^2}\right) \right].
\end{eqnarray}
%------------------------------------------------------------------------------------------------------
The OMEs obey the sum rule (\ref{eq:SR}) in both cases because
%------------------------------------------------------------------------------------------------------
\begin{eqnarray}
\frac{\hat{\gamma}_{qg}^{(0)}(N=2)}{2} - \beta_{0,Q} = 0
\end{eqnarray}
%------------------------------------------------------------------------------------------------------
holds. The two-mass contributions (\ref{eq:tm1}, \ref{eq:tm2}) at NLO are the first terms of this kind 
emerging and are the same in both schemes. 
The corresponding values of the heavy quark masses in the 
$\overline{\sf MS}$ scheme are $\overline{m}_c = 1.24~\GeV$ and $\overline{m}_b = 4.18~\GeV$.
The numerical integrals emerging in the present calculation have been performed using the code {\tt AIND} 
\cite{AIND} and the harmonic polylogarithms
have been evaluated using the package {\tt hplog} \cite{Gehrmann:2001pz}, cf. also \cite{Ablinger:2017tqs}.
The additional two-mass terms described in the present paper are of logarithmic order and are therefore of 
comparable size to the terms appearing in the single mass case.

The VFNS is used in many applications, cf. e.g. \cite{Accardi:2016ndt}, and has even been advocated by the 
{\tt pdf4lhc} recommendation \cite{Alekhin:2011sk} for use. Its correct use is also of importance for all processes 
at hadron colliders, such as the Tevatron and the LHC, with charm and bottom quarks in the initial state.
The corresponding former parameterizations have to be changed according to the relations 
(\ref{eq:FIS}--\ref{eq:ma2VNS})
as a consequence. Furthermore, in precision measurements of the strong coupling constant $\alpha_s(M_Z^2)$ 
\cite{ALPHA}, the charm and bottom quark masses and the parton distribution functions, if working in the VFNS, 
the correct relations have to be applied.

Since in QCD fits the structure function $F_2(x,Q^2)$ plays an important role we present the
two-mass contributions to this observable for pure virtual photon exchange. It is given by
%------------------------------------------------------------------------------------------
\begin{eqnarray}
F_2^{{\rm 2-mass},(2)}(x,Q^2) = \frac{32}{3} T_F^2 a_s^2(Q^2) x 
\ln\left(\frac{Q^2}{m_c^2}\right)
\ln\left(\frac{Q^2}{m_b^2}\right)
\int_x^1 \frac{dy}{y} [y^2 + (1-y)^2] 
G\left(\frac{x}{y},Q^2\right),
\end{eqnarray}
%------------------------------------------------------------------------------------------
choosing the renormalization and factorization scale $\mu^2 = Q^2$. We mention that for the 
{\it inclusive} heavy flavor contribution to $F_2(x,Q^2)$ also the single heavy quark contributions of
Ref.~\cite{Bierenbaum:2009zt} have to be added working in the ${\overline{\rm MS}}$ scheme for the coupling
constant renormalization, which are sometimes missing in the codes following Ref.~\cite{Buza:1996wv}. 
These contributions stem from massless final states with virtual heavy quark corrections.

We add a word of caution on the use of parton distributions in the VFNS, as e.g. in the representation 
(\ref{eq:FI}--\ref{eq:ma2VNS}). In assembling any observable up to a certain order in the coupling, $a_s$, e.g. 
$l$, 
the factorization theorem\footnote{See Ref.~\cite{Behring:2014eya}, Eqs.~(11, 19--27).}  leads to the 
cancellation of the 
factorization scale $\mu^2_F = \mu^2$. However, the required matching is not global. 0th order Wilson 
coefficients match 
to $l$th order OMEs and contributions to parton distributions, 1st order Wilson coefficients to $(l-1)$st 
order OMEs and PDFs, etc. If this matching is disregarded, a corresponding $\mu$-dependence is implied, 
which in principle can be thoroughly avoided, cf. e.g.~Ref.~\cite{Blumlein:2000wh}.

\begin{figure}[H]\centering
\includegraphics[width=0.7\textwidth]{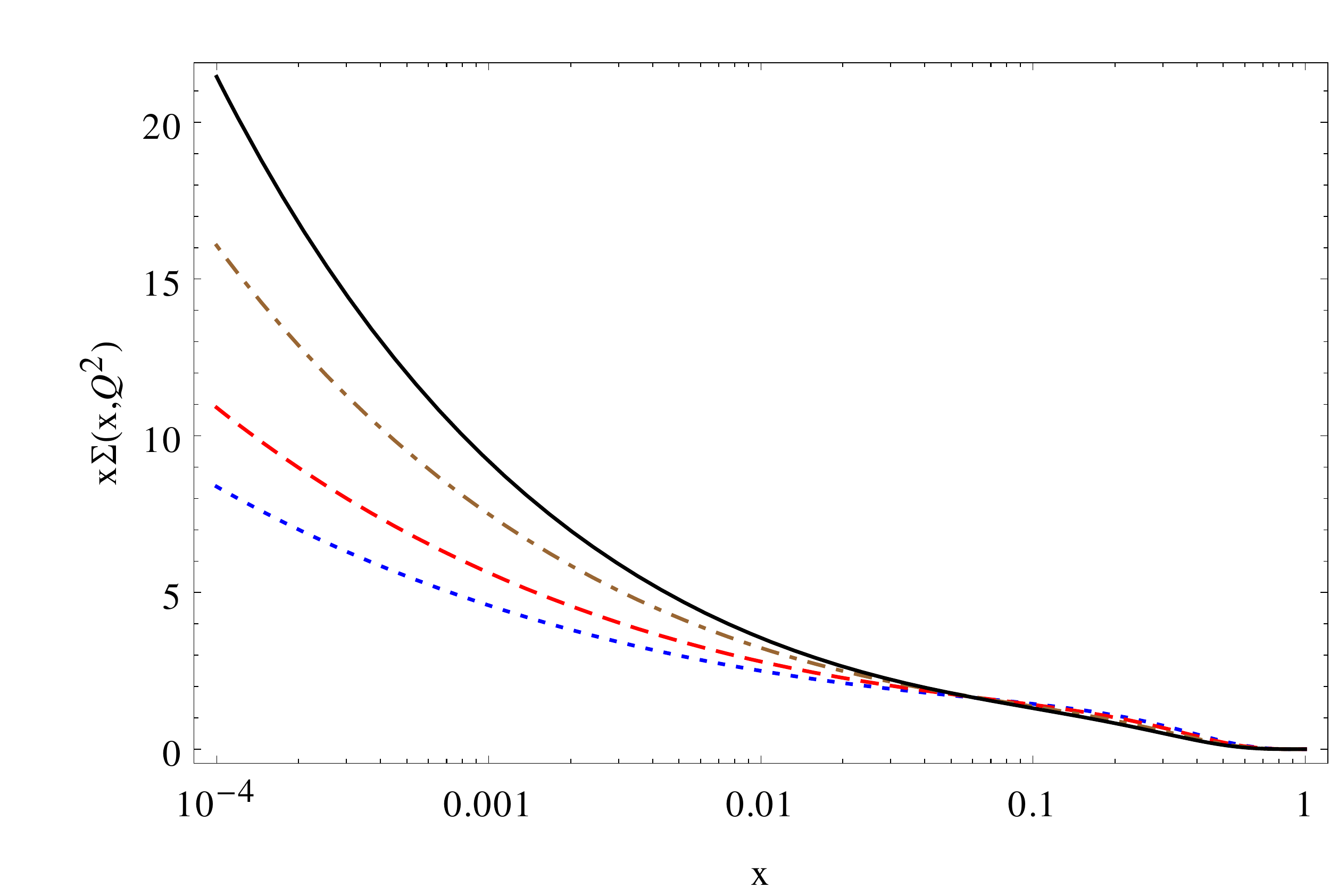}
\caption[]{\label{fig:SIG} \sf
The singlet distribution $x\Sigma(x,Q^2)$  as a function of $x$ and $Q^2$ using the parton 
distribution functions
\cite{Alekhin:2017kpj}. 
Dotted line: $Q^2 = 30~\GeV^2$;
dashed line: $Q^2 = 100~\GeV^2$;
dash-dotted line: $Q^2 = 1000~\GeV^2$;
full line: $Q^2 = 10000~\GeV^2$.
}
\end{figure}

\begin{figure}[H]\centering
\includegraphics[width=0.7\textwidth]{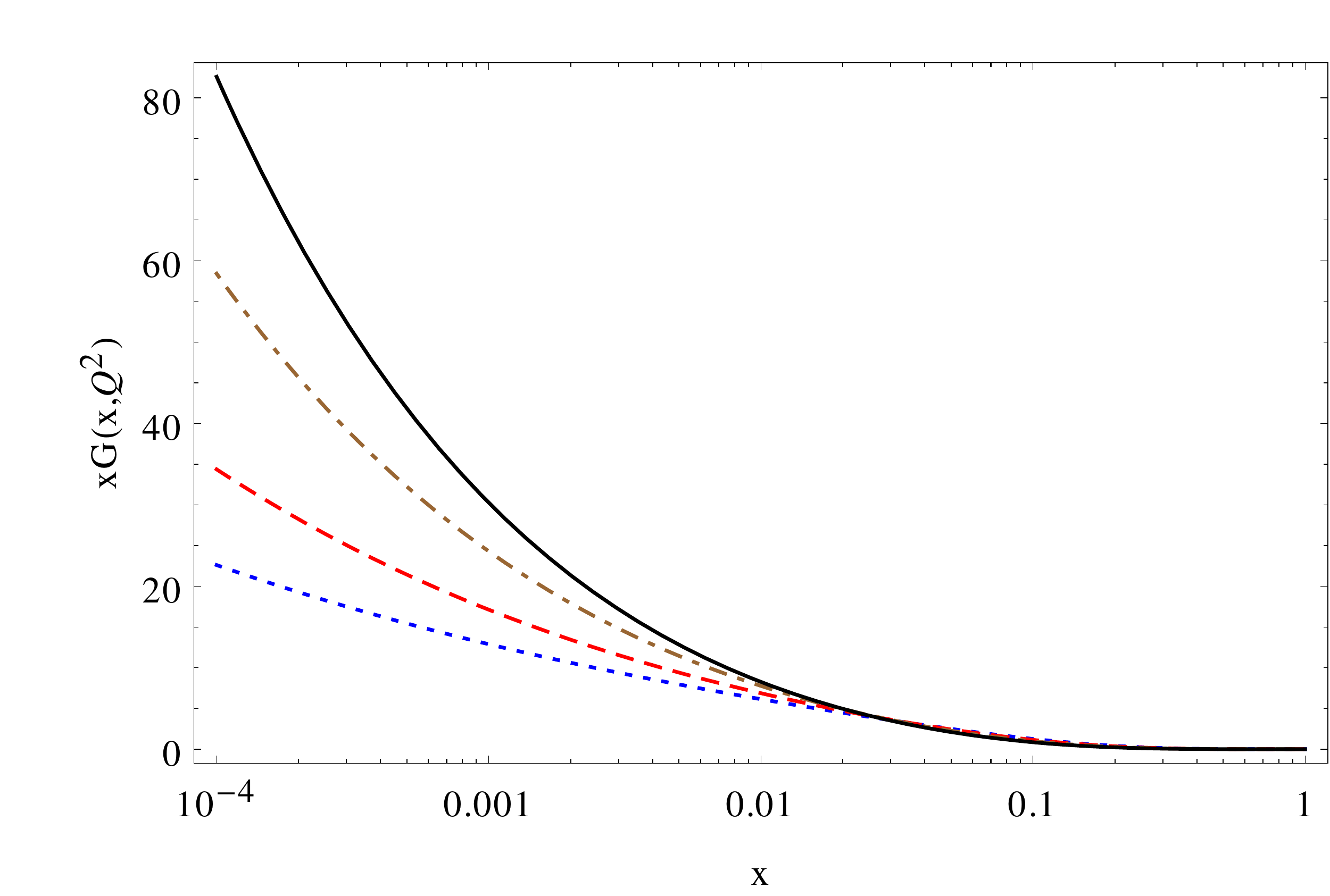}
\caption[]{\label{fig:GL0} \sf
The gluon distribution $xG(x,Q^2)$ as a function of $x$ and $Q^2$ using the parton distribution functions
\cite{Alekhin:2017kpj}.
Dotted line: $Q^2 = 30~\GeV^2$;
dashed line: $Q^2 = 100~\GeV^2$;
dash-dotted line: $Q^2 = 1000~\GeV^2$;
full line: $Q^2 = 10000~\GeV^2$.
}
\end{figure}

In the following numerical illustration, we refer to the parton distribution functions at 
NNLO \cite{Alekhin:2017kpj}\footnote{Very recently, a NLO variant of this fit has been presented in 
\cite{Alekhin:2018pai}.}, implemented in 
{\tt LHAPDF} \cite{Buckley:2014ana}. The 
flavor singlet and gluon momentum distributions for $N_F=3$ are depicted in Figures~\ref{fig:SIG} and \ref{fig:GL0}
as functions of the Bjorken variable $x$ and the virtuality $Q^2$ for reference.
%------------------------------------------------------------------------------------------------------------------------
%------------------------------------------------------------------------------------------------------------------------

\begin{figure}[H]\centering
\includegraphics[width=0.7\textwidth]{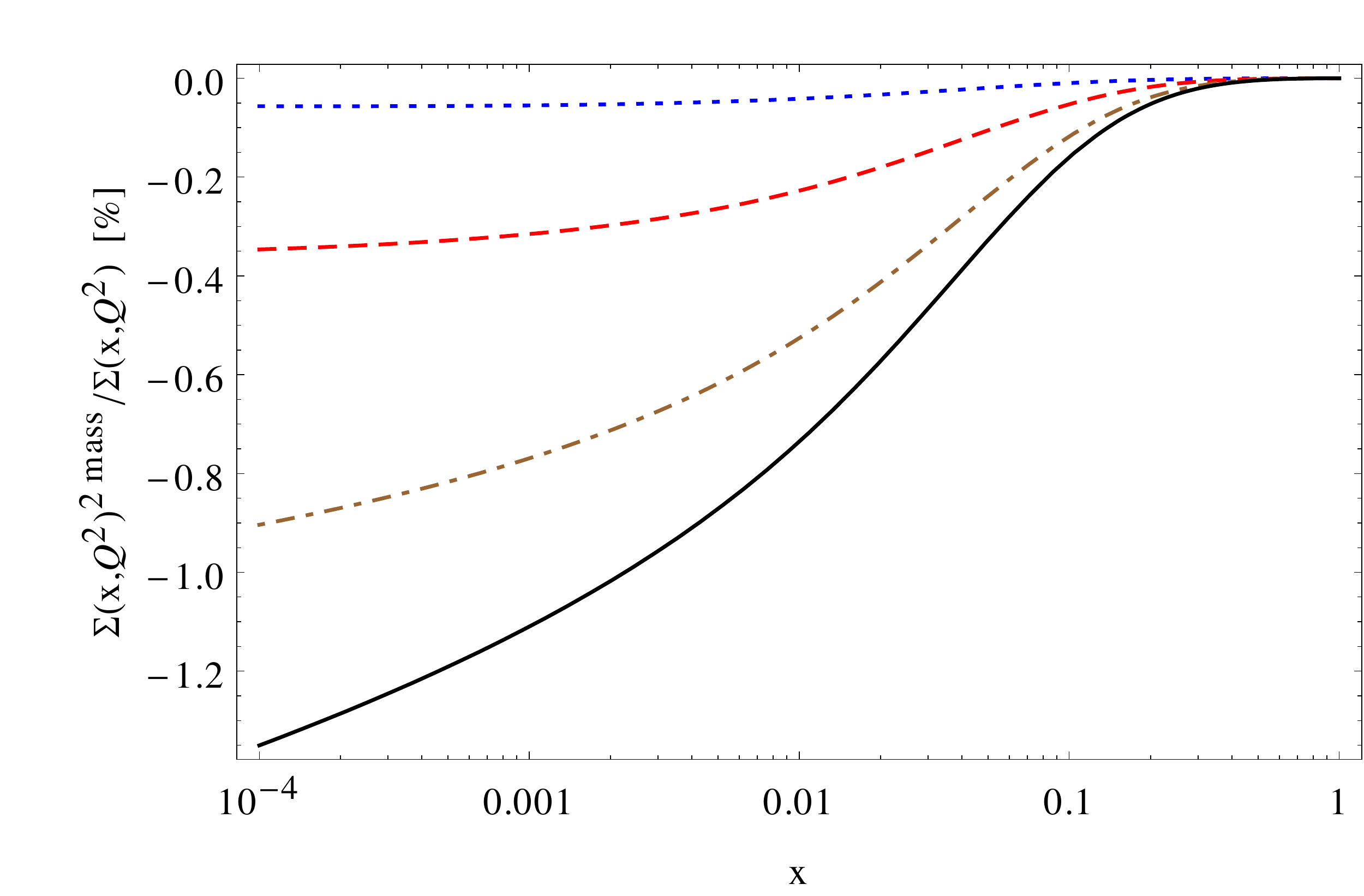}
\caption[]{\label{fig:SIGR} \sf
The ratio of the two-mass contribution to the 
singlet distribution and the complete singlet distribution at $O(a_s^2)$, Eq.~(\ref{eq:FIS}), in \%, as a 
function of $x$ and $Q^2$, 
using the parton distribution functions
\cite{Alekhin:2017kpj} and $m_c =1.59~\GeV$ \cite{Alekhin:2012vu}, $m_b = 4.78~\GeV$ \cite{Agashe:2014kda}.
Dotted line: $Q^2 = 30~\GeV^2$;
dashed line: $Q^2 = 100~\GeV^2$;
dash-dotted line: $Q^2 = 1000~\GeV^2$;
full line: $Q^2 = 10000~\GeV^2$.
\label{FIG:SIGRAT}
}
\end{figure}

\begin{figure}[H]\centering
\includegraphics[width=0.7\textwidth]{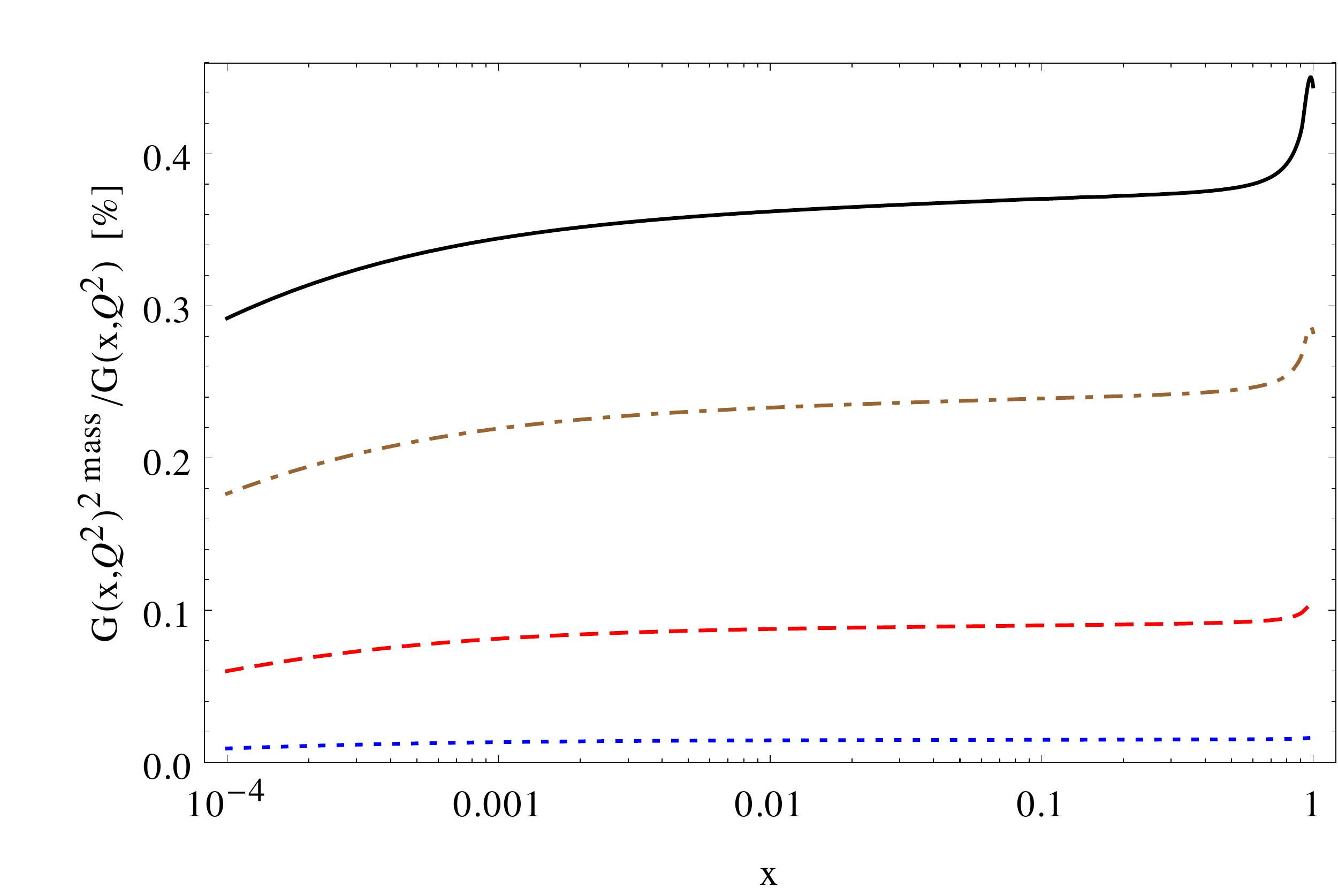}
\caption[]{\label{fig:GLUR} \sf
The ratio of the two-mass contribution to the 
gluon distribution and the complete gluon distribution at $O(a_s^2)$, Eq.~(\ref{eq:GNF2}), in \%, as a function 
of $x$ and $Q^2$ 
using the parton distribution functions
\cite{Alekhin:2017kpj} and $m_c =1.59~\GeV$ \cite{Alekhin:2012vu}, $m_b = 4.78~\GeV$ \cite{Agashe:2014kda}.
Dotted line: $Q^2 = 30~\GeV^2$;
dashed line: $Q^2 = 100~\GeV^2$;
dash-dotted line: $Q^2 = 1000~\GeV^2$;
full line: $Q^2 = 10000~\GeV^2$.
\label{FIG:GLRAT}
}
\end{figure}

\begin{figure}[H]\centering
\includegraphics[width=0.7\textwidth]{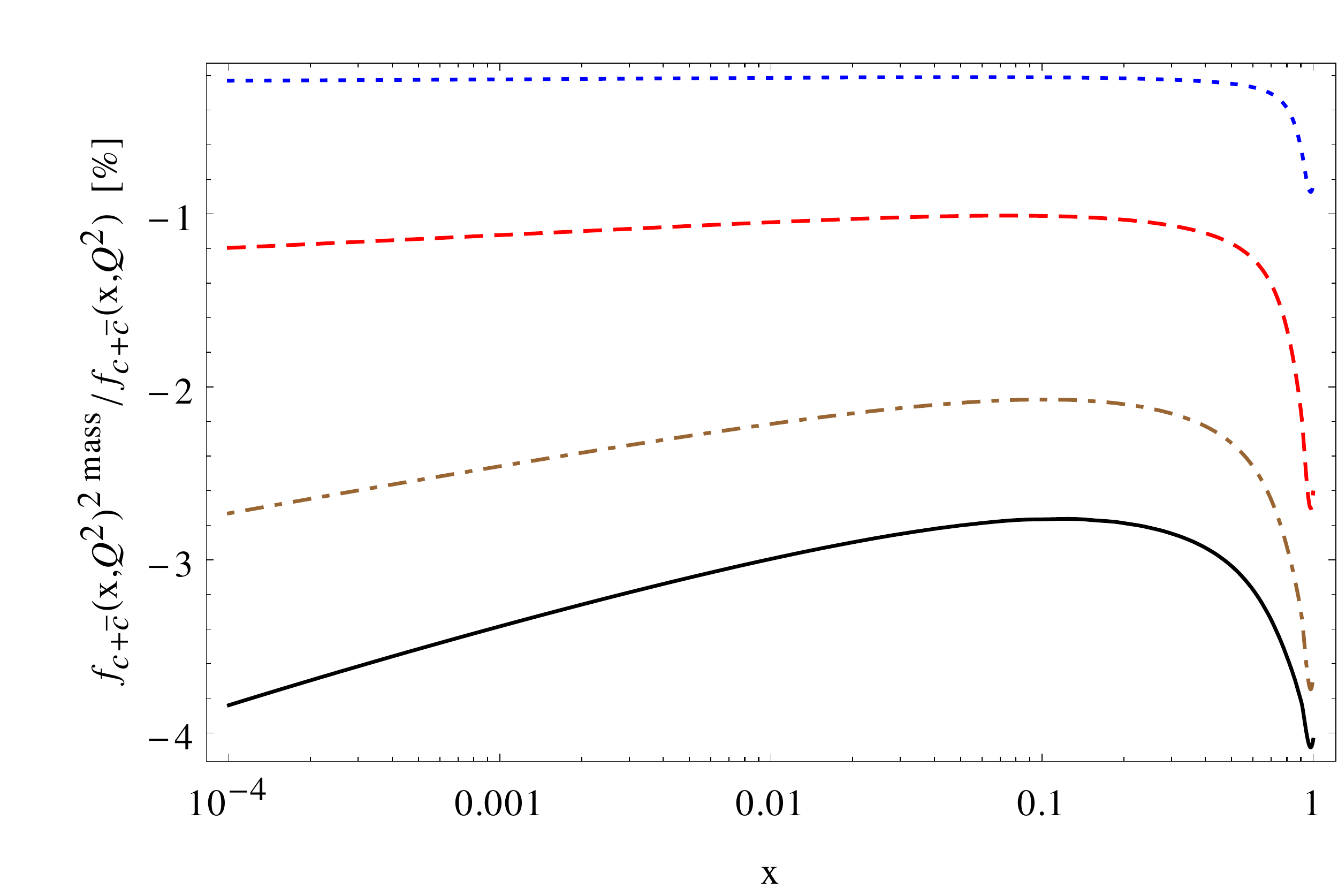}
\caption[]{\label{fig:HQC} \sf
The ratio of the two-mass contribution to the 
charm distribution and the complete charm distribution at $O(a_s^2)$, Eq.~(\ref{eq:cc}), as a function of $x$ 
and 
$Q^2$ 
using the parton distribution functions
\cite{Alekhin:2017kpj} and $m_c =1.59~\GeV$ \cite{Alekhin:2012vu}, $m_b = 4.78~\GeV$ \cite{Agashe:2014kda}.
Dotted line: $Q^2 = 30~\GeV^2$;
dashed line: $Q^2 = 100~\GeV^2$;
dash-dotted line: $Q^2 = 1000~\GeV^2$;
full line: $Q^2 = 10000~\GeV^2$.
\label{FIG:CRAT}
}
\end{figure}

\begin{figure}[H]\centering
\includegraphics[width=0.7\textwidth]{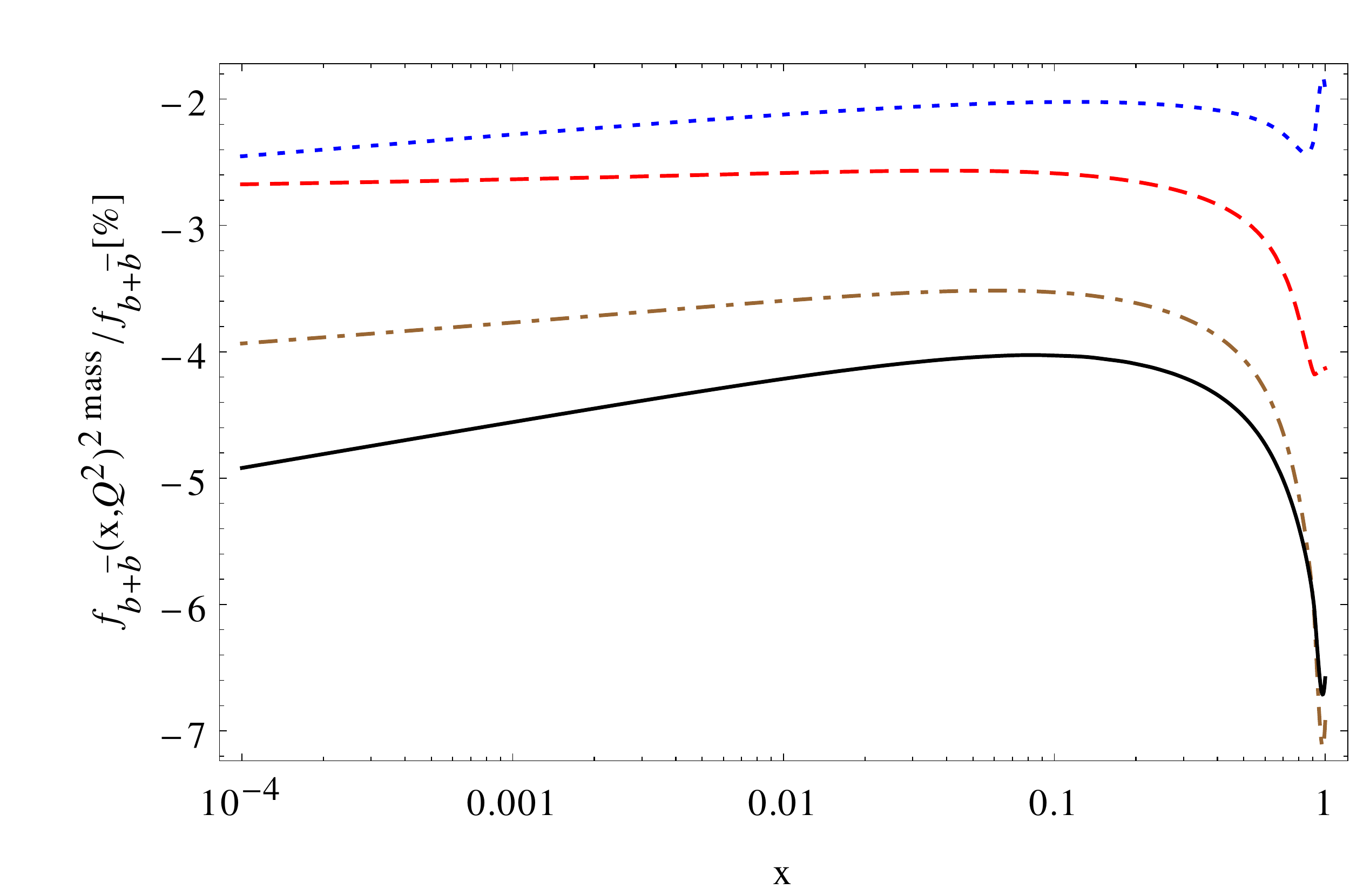}
\caption[]{\label{fig:HQB} \sf
The ratio of the two-mass contribution to the 
bottom distribution and the complete bottom distribution at $O(a_s^2)$, Eq.~(\ref{eq:ma2VNS}) as a function of 
$x$ and $Q^2$ 
using the parton distribution functions
\cite{Alekhin:2017kpj} and $m_c =1.59~\GeV$ \cite{Alekhin:2012vu}, $m_b = 4.78~\GeV$ \cite{Agashe:2014kda}.
Dotted line: $Q^2 = 50~\GeV^2$;
dashed line: $Q^2 = 100~\GeV^2$;
dash-dotted line: $Q^2 = 1000~\GeV^2$;
full line: $Q^2 = 10000~\GeV^2$.
\label{FIG:BRAT}
}
\end{figure}

\noindent
In Figures~\ref{fig:SIGR}--\ref{fig:HQB} we show the ratios of the two-mass contributions to the total rate 
for the flavor singlet, gluon, charm and bottom contributions up to $O(a_s^2)$ as functions of $x$ and $Q^2$ 
according to (\ref{eq:FIS}--\ref{eq:ma2VNS}) in the on-mass shell scheme, setting $\mu^2 = Q^2$. 
We use the OMEs calculated in Refs.~\cite{Behring:2014eya} in the $\overline{\sf MS}$ scheme for the 
strong coupling constant and the parton distribution functions, while the heavy quark masses are given in the 
on-mass shell scheme. To put the numerical effects into the perspective of later NNLO corrections we will present 
the illustration choosing the NNLO values for $a_s$, the heavy quark masses with $m_c = 1.59~\GeV$ and $m_b = 
4.78~\GeV$, cf.~\cite{Alekhin:2012vu,Agashe:2014kda}.
%------------------------------------------------------------------------------------------------------------------------
%------------------------------------------------------------------------------------------------------------------------

%------------------------------------------------------------------------------------------------------------------------
%------------------------------------------------------------------------------------------------------------------------

\noindent
The two-mass corrections to the singlet distribution 
in Figure~\ref{FIG:SIGRAT}, are negative and their relative 
contribution varies between 
$\sim 0.06 \%$ at $Q^2 = 30~\GeV^2$ to $\sim 1.4 \%$ at $Q^2 = 10000~\GeV^2$ at $x = 10^{-4}$ diminishing in 
modulus towards larger values of $x$.
%------------------------------------------------------------------------------------------------------------------------
%------------------------------------------------------------------------------------------------------------------------

\noindent
The relative contribution of the NLO 2-mass term to the gluon distribution for $N_F+2$ flavors, shown in 
Figure~\ref{FIG:GLRAT}, is positive 
and shows a slightly rising behaviour in $x$ and grows with $\mu^2$ from values of $\sim 0.01\%$ at $\mu^2 = 
30~\GeV^2$ to  $\sim 0.4 \%$ at $\mu^2 = 10000~\GeV^2$. Here the positive correction balances the negative
quarkonic corrections for the singlet and the heavy quark contributions.
%------------------------------------------------------------------------------------------------------------------------
%------------------------------------------------------------------------------------------------------------------------
%------------------------------------------------------------------------------------------------------------------------
%------------------------------------------------------------------------------------------------------------------------

\noindent
Figures~\ref{FIG:CRAT} and \ref{FIG:BRAT} show the relative two-mass corrections for  the charm and bottom 
quark distributions. They are both negative and are slightly rising in the low $x$ region and become larger
in size for large values of $x$, where the distributions themselves are very small, however. For charm the 
largest
corrections at $x = 10^{-4}$ vary between $\sim -0.2 \%$ ($Q^2 = 30~\GeV^2$) and $\sim -3.8 \%$ ($Q^2 = 
10000~\GeV^2$) and for bottom the corresponding values are  $\sim -2.5 \%$ ($Q^2 = 50~\GeV^2$) and $\sim -4.9 
\%$ ($Q^2 = 10000~\GeV^2$). Here we have chosen a somewhat larger lowest scale because of the heavier quark mass.
Comparing the different relative corrections, the largest are those for the bottom 
distribution, as expected, cf.~(\ref{eq:EST}). Similar numerical results are obtained using other sets of parton 
distributions, as e.g. the GRV98 distributions \cite{Gluck:1998xa}.

One may sometimes resum, at least to leading order, mass logarithms into the parton densities or the coupling 
constant or into both. In doing this, 
one changes the scheme, however, from the $\overline{\sf MS}$-scheme, in which the comparison of the different 
fitted coupling constants and/or the parton distribution functions for different analyses is performed under 
well defined conditions, to another new scheme. The latter now depends in many places on the chosen value
of the quark masses and changes with them. As a consequence, the corresponding coupling constants and parton 
densities cannot be compared at all anymore. This has to be considered in precision measurements
of the strong coupling constant, of the heavy quark masses, and the parton distribution functions.

\vspace*{5mm}
\noindent
{\bf Acknowledgment.} We thank A.~Behring for discussions. This work was supported in part by the
Austrian Science Fund (FWF) grant SFB F50 (F5009-N15).

\newpage
{\small
%-------------------------------------------------------------------------------------------------------------------------
}
%------------------------------------------------------------------------
\end{document}